\documentclass[reprint,aip,superscriptaddress,amsmath,longbibliography]{revtex4-2}
\usepackage{graphicx,txfonts}
\usepackage[dvipdfm,colorlinks,breaklinks,linkcolor=blue,citecolor=blue,anchorcolor=blue,urlcolor=blue]{hyperref}
\usepackage{pifont}
\usepackage{subfigure}
\usepackage{overpic}

\begin{document}

\title{Organization of cooperation in fractal structures}

\author{Dan Peng}
\affiliation{Department of Network and New Media, Anhui University, Hefei 230601, P. R. China}

\author{Ming Li}
\email{minglichn@ustc.edu.cn}
\affiliation{Department of Thermal Science and Energy Engineering, University of Science and Technology of China, Hefei 230026, P. R. China}

\date{\today}

\begin{abstract}
It is known that the small-world structure constitutes sufficient conditions to sustain cooperation and thus enhances cooperation. On the contrary, the network with a very long average distance is usually thought of as suppressing the emergence of the cooperation.  In this paper we show that the fractal structure, of which the average distance is very long, does not always play a negative role in the organization of cooperation. Compared to regular networks, the fractal structure might even facilitate the emergence of cooperation. This mainly depends on the existence of locally compact clusters. The sparse inter-connection between these clusters constructs an asymmetric barrier that the defection strategy is almost impossible to cross, but the cooperation strategy has a not too small chance. More generally, the network need not to be a standard fractal, as long as such structures exist. In turn, when this typical structure is absent, the fractal structure will also suppress the emergence of the cooperation, such as the fractal configuration obtained by diluting a random tree-like network. Our findings also clarify some contradictions in the previous studies, and suggest that both removing and inserting links from/into a regular network can enhance cooperation.
\end{abstract}

\maketitle

\section{Introduction}

In the study of complex systems, a long-standing question is how the collective behavior of interacting individuals is influenced by the topology of those interactions. Inspired by network science, the interaction topologies extracted from real networked systems have been widely studied in the last two decades \cite{Albert2002,Dorogovtsev2008}. Two typical examples are the structures with scale-free property and small-world effect.

Mathematically, the scale-free property translates into a power-law degree distribution of a network, where a node's degree is the number of nodes that are adjacent to the node. Obviously, this formula allows the existence of very high degree nodes, called hubs of networks, rather than the comparable degrees in Erd\"os-R\'enyi (ER) networks or regular networks \cite{Barabasi2016}. As a result, the dynamic on such a network bears a strong heterogeneity induced by the existence of hubs \cite{Dorogovtsev2008}. Specifically, the change of a hub's state can affect a large number of nodes, while the influences of periphery nodes are limited to their few neighbors.

It is conceivable that the average distance between nodes is also an important characteristic of a topology structure \cite{Barabasi2016}. A short average distance can facilitate communications between nodes, while a long one lengthens the response time for communications. If the average distance $l$ grows proportionally to the logarithm of the network size $N$, i.e., $l\sim \log N$, it just refers to the small-world effect \cite{Watts1998}. In general, the small-world effect can be constructed by inserting long-range connections into a regular network \cite{Watts1998}. Besides this, there is another typical topology called fractal \cite{Bunde2012}, showing an average distance in a power-law manner with the system size. Obviously, for the same network size, the average distance in this form is much longer than that of the logarithmic form. In the study of network dynamics, a long distance is usually considered to be not favorable for the spreading process \cite{Rong2007,Li2009,Zhang2008,Wang2017}. As a contrast, the small-world network can reduce the threshold for an epidemic outburst \cite{Moore2000,Santos2005}, yield the most effective information spreading \cite{Lue2011}, and promote cooperation \cite{Santos2005a,Santos2005b,Barrat2007,Yun2011}.

The fractal structure is not the only topology that has an average distance in the form of a power-law with network size. The regular lattice also has such a form of average distance. However, the fractal can further exhibit similar patterns at increasingly smaller scales, called self similarity \cite{Bunde2012}, which distinguishes the dynamics on fractal structures from that of regular lattices. For instance, in an evolutionary game model \cite{Wang2012,Wang2012a}, Wang et al. found that compared to the system far away from the percolation threshold, which can be seen as more like a regular network, the system at the percolation threshold, where the system shows fractal structure \cite{Stauffer1991}, can optimally promote cooperation. Note that in these two works, the fractal was not mentioned to explain the finding.

Moreover, the fractal structure does not always enhance cooperation as shown in Ref.\cite{Wang2012,Wang2012a}. Tang et al. found that an ER network requires a relatively dense connectivity to achieve an optimal cooperation \cite{Tang2006}, however, for average degree $1$, of which ER network (the giant component) shows a fractal structure, no cooperators can survive. This indicates that the effect of fractal structure on the cooperation evolution is more than just a result of a long average distance.

The previous studies also suggested that a long average distance can play different roles in different models. For example, some works have shown that a long average distance could reduce cooperation \cite{Santos2005a,Santos2005b,Yun2011}, while some others showed that a reasonable number of long-range connections can enhance cooperation \cite{Perc2006,Vilone2011,Perc2008}. Moreover, some studies of coevolutionary games \cite{Perc2010}, such as inertia \cite{Liu2010}, reconnection\cite{Zhang2011}, catastrophic failures \cite{Wang2011}, and conditional participation \cite{Li2013}, indicated that diluting the connection of a network, which leads to a long average distance, could facilitate the emergence of cooperation.

In this paper we will employ the prisoner's dilemma game (PDG) to study the organization of cooperation in fractal structures and show that the fractal has its own structural feature to influence the evolution of cooperation, which cannot be simply featured by the long average distance. In the followings, we will first introduce the model used in this study, and then the results on two different fractal structures will be discussed.

\section{Cooperation evolution}

Initially, cooperators (C) and defectors (D) are randomly and uniformly distributed over the network. In accordance with the standard definition of the PDG, we set the payoff of an individual encountering with (adjacent to) a D is $0$, in contrast with encountering with a C, a C and a D respectively gain payoff $1$ and $b$ ($b>1$). To represent the cooperation evolution, at each time step a randomly chosen individual $x$ tries to enforce its strategy (C or D) on a randomly chosen neighbor $y$ with the probability
\begin{equation}
w_{x\to y}=\frac{1}{1+e^{-(p_x-p_y)/\kappa}},  \label{updr}
\end{equation}
where $p_x$ ($p_y$) is the total payoff of individual $x$ ($y$) gaining from all its neighbors, and $\kappa$ determines the level of uncertainty by strategy adoptions \cite{Szabo1998}. Without loss of generality we set $\kappa=0.1$ in this paper. Note that when an individual's strategy changed, all the payoffs gaining from this individual need to be recalculated. For convenience, we rescale the time step by the system size $N$, thus one unit contains $N$ time steps, called a Monte Carlo step.

\section{Hierarchical lattice}

\begin{figure}
\centering
\scalebox{0.3}{\includegraphics{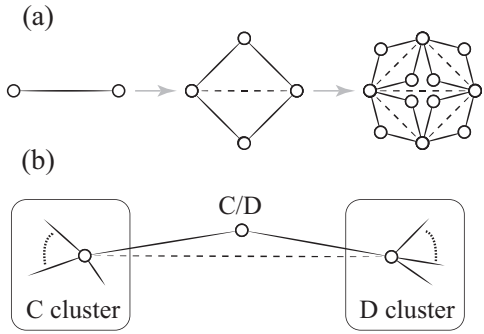}}
\caption{Schematics of the hierarchical lattice. (a) As indicated by the grey arrows, a normal link will grow to be a diamond-type lattice when the generation of hierarchical lattice increases. Meanwhile, the original link (indicated by dashed line) will be preserved as a long-range link with probability $p$. (b) A typical local structure in the hierarchical lattice. That is two large degree nodes connected through two degree-$2$ nodes (only one such node is shown here), and with probability $p$ there is also a long-range link (indicated by dashed line) between the two nodes.} \label{fig1}
\end{figure}

To study the effects of the fractal structure on the evolution of cooperation, we first consider a hierarchical lattice, which can demonstrate both fractal and small-world characteristics by applying different connection probability $p$ \cite{Hinczewski2006}. This network model adopts a growth mechanism from a single link, called generation $n=0$. The generation $n+1$ is constructed by expanding each normal link (those are not long-range links) of generation $n$ to be a diamond-type lattice, and the original link is preserved as a long-range link with probability $p$, see Fig.\ref{fig1} (a). Note that long-range links do not participate in this growth process.

For $p=1$, the network exhibits the typical small-world property with a high clustering coefficient. Due to the absence of long-range links, the case $p=0$ is a large-world lattice exhibiting an average distance $l\sim N^{1/2}$ \cite{Hinczewski2006}. As shown in Fig.\ref{fig1} (b), a typical local structure of this network is that two large degree nodes are bridged by two degree-$2$ nodes, meanwhile, there exists a link (long-rang link) between the two nodes with probability $p$. In fact, there is often a compact cluster on the basis of a large degree node. Thus, the case of $p\to0$ just shows a typical structure composed of compact clusters and sparse inter-connections between them. As will be discussed later, this typical structure plays a key role in the evolution of cooperation on fractal structures.

\begin{figure}
\centering
\scalebox{0.3}{\includegraphics{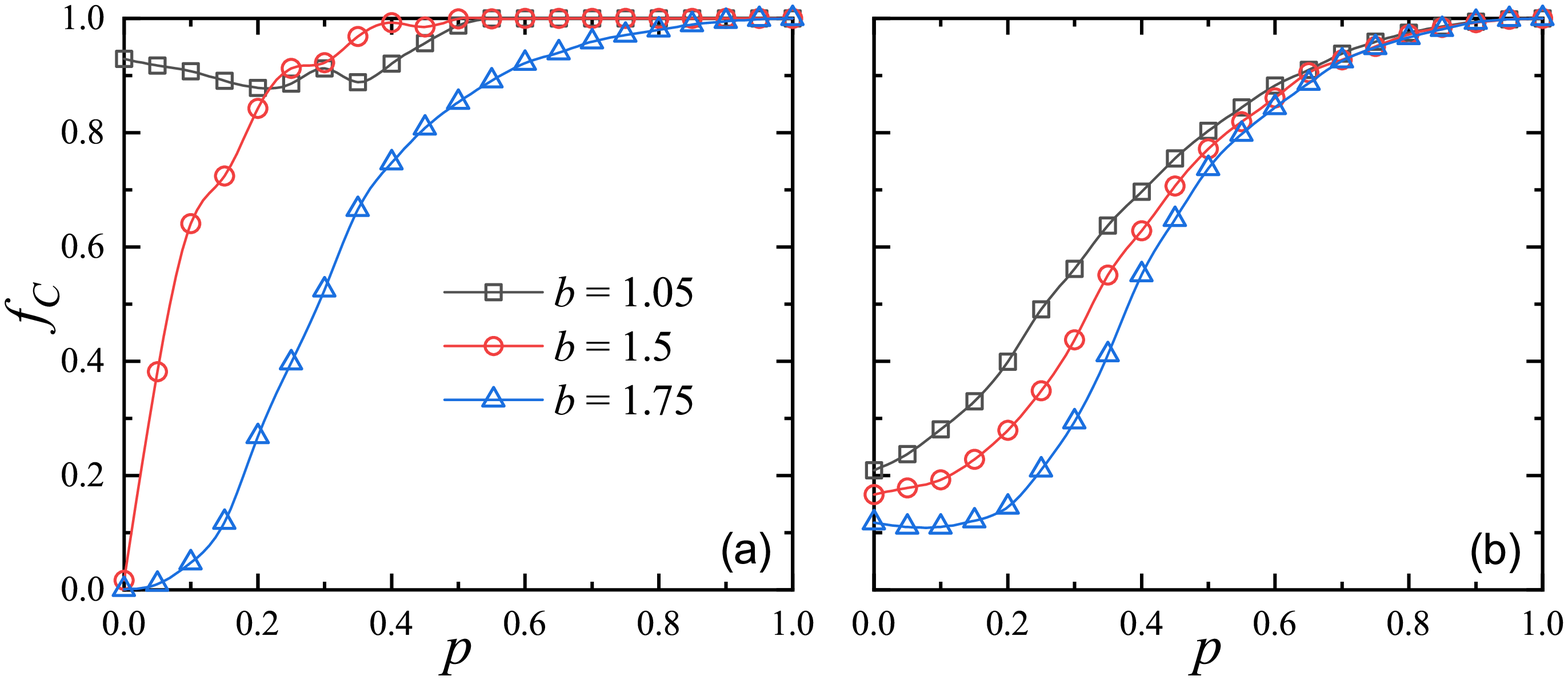}}
\caption{(a) Density of cooperators $f_C$ as a function of connection probability $p$ of the hierarchical network. Here, the hierarchical network have $n=7$ generations, that is $N=10924$. For $p=0$, the long-range link is absent, and the system shows a fractal structure. With the increasing of $p$, long-range links are inserted, and the average distance of the system decreases. The system demonstrates the small-world property. (b) The case that an individual can only enforce its strategy to the one with a payoff smaller than its own. All the other settings are the same as that used in (a). } \label{fig2}
\end{figure}

It is widely acknowledged that the small-world effect can significantly enhance the cooperation \cite{Santos2005a,Santos2005b,Gomez-Gardenes2007}. To demonstrate this, in Fig.\ref{fig2} (a) we show the cooperator density $f_C$ in hierarchical networks under three relatively large $b$, for which the regular lattice, such as the square lattice (degree $4$ for each node) and the honeycomb lattice (degree $3$ for each node) \footnote{The hierarchical lattice has average degree $3+p$ in the infinite lattice limit \cite{Hinczewski2006}}, gives the full D state \cite{Szabo2005}. We can find that due to the small-world effect, a large $p$ always corresponds to a high cooperator density $f_C$. However, compared to regular lattices, which give the full D state for the parameters used in Fig.\ref{fig2} (a) \cite{Szabo2005}, the fractal structure ($p\to0$) is also in favour of the emergence of the cooperation. This suggests that the hierarchical lattice with large $p$ (small-world effect) does not have a distinctly unique advantage in the emergence of cooperation. Without the long-range link, the typical structure of fractal structures can also constitute sufficient conditions to sustain cooperation, and even induces a nearly full C state (see the case $b=1.05$ in Fig.\ref{fig2} (a)). To understand how that works, we next check the spreading of the two strategies on such a fractal structure.

Specially, assuming that the two compact clusters at the two ends of the degree-$2$ node are a C cluster and a D cluster, respectively (see Fig.\ref{fig1} (b)). For fractal structures, the long-range link is absent, so that the growths of the two clusters have to go through the degree-$2$ node. If the degree-$2$ node is a D, it can gain a payoff $b$ from the C neighbor in the C cluster. However, this payoff is much smaller than the payoff of its C neighbor, labeled as $m$ for convenience, since it can gain mutual benefit from the C cluster. From Eq.(\ref{updr}), we can know that the probability that D can invade into the C cluster through the degree-$2$ node is $w_{b,m}=[1+e^{(m-b)/\kappa}]^{-1}$, which is very small for large $m$. This means that the C cluster in such a structure is much more stable than that in a regular lattice.

In turn, if the degree-$2$ node is a C, it gains payoff $1$ from its C neighbor in the C cluster, while its D neighbor in the D cluster gains payoff $b$ from it. Since $b>1$, the degree-$2$ node will change its strategy to be a D with a relatively large probability, then this reduces to the former case. Nevertheless, since the value $b$ of interest in the study is often not much larger than $1$, the update rule Eq.(\ref{updr}) also allows a not too small chance $w_{1,b}=[1+e^{(b-1)/\kappa}]^{-1}$ that the degree-$2$ node of strategy C can enforce its strategy to its D neighbor in the D cluster.

In conclusion, the probability $w_{1,b}$ that C can invade into the D cluster could be much larger than the probability $w_{b,m}$ that D can invade into the C cluster. Taking $b=1.05$ and $m=3$ as an example, we have $w_{1,b}\approx0.38$ and $w_{b,m}\approx3.4\times10^{-9}$. This means that the structure shown in Fig.\ref{fig1} (b) (without the long-range link) constructs an asymmetric barrier for the growth of C and D clusters, for which D strategy is almost impossible to cross, while C strategy has a not too small chance. In other words, the C cluster is more stable than the D cluster. This explains why for small $p$ the hierarchical lattice can also show a very high cooperation level. Moreover, a very large $b$ will break the asymmetry, and this structure would no longer facilitate cooperation, see the cases $b=1.5$ and $b=1.75$ of Fig.\ref{fig2} (a).

In fact, this asymmetric barrier comes from that an individual has a small chance to enforce its strategy to the one with a larger payoff. If such a strategy update is not allowed, namely, individuals can only enforce their strategies to the ones with payoffs smaller than theirs, this structure will no longer facilitate the emergence of the cooperation. The corresponding results are shown in Fig.\ref{fig2} (b). This further confirms the above analysis, and is also consistent with the result found in Ref.\cite{Yun2011}.

\begin{figure}
\centering
\scalebox{0.3}{\includegraphics{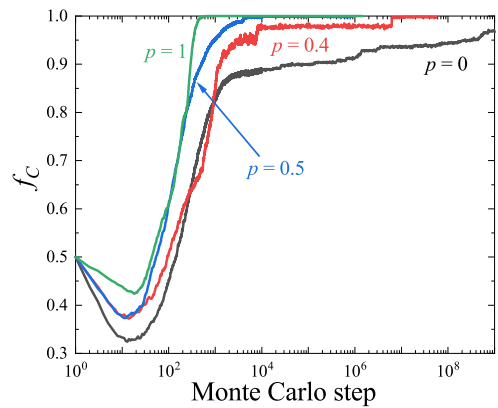}}
\caption{The dynamical temporal behaviors of PDG in hierarchical networks with different connection probability $p$. Here, the parameters are set as $N=10924$ and $b=1.05$. It can be observed that the convergence time of the system increases as $p$ decreases. For $p=0$, the convergence even cannot be achieved in an acceptable time. Moreover, we can find some step-like growth in these curves, each of which corresponds to a complete conversion of a D cluster.} \label{fig3}
\end{figure}

As a second consequence, the fractal structure can also stretch the convergence time of the system as shown in Fig.\ref{fig3}. This is mainly due to two reasons. First, the long average distance obviously extends the spreading process. Second, due to the sparse inter-connections between the compact clusters of fractal structures, a successful invasion into D cluster might need a larger number of attempts and thus lead to a very large convergence time. An immediate consequence of this is that the long-term growth of the cooperator density $f_C$ has many abrupt changes, where a jump corresponds to a complete conversion of a D cluster, see Fig.\ref{fig3}. It should be noted that even for this case, the case of $p=0$ gives a non-vanished $f_C$ rather than the full D state.

Generally speaking, the spreading of strategies is cramped by the sparse inter-connections between different clusters, but not completely blocked. This increases the convergence time and ultimately favours the emergence of cooperation. Note that for small $p$, the convergence time is so large that there is no uniform criterion for the convergence of the system. For some cases, the convergence even cannot be achieved in an acceptable time, see case $p=0$ in Fig.\ref{fig3}. From this perspective, we suspect that for small $b$, such as the case $b=1.05$ shown in Fig.\ref{fig2} (a), a large enough system will also give a full C state for small $p$. However, due to the high time complexity, we have no firm numerical evidence to support this.

It is also worth pointing out that the compact cluster mentioned here is not equivalent to the community structure, although a large convergence time was also reported in the network with community structures \cite{Gianetto2015}. In general, the community is a macrostructure scaling in the order of the system size, while the compact cluster mentioned here is a microstructure containing only a small amount of nodes. In a community different strategies could coexist, while there is often only one strategy in a small compact cluster.

\section{Networks near the percolation threshold}

Excluding the small-word effect, hubs also exist in the hierarchical lattice, which has also been proved to be extremely important in the evolution of the cooperation \cite{Santos2005a,Santos2005b,Tang2006,Gomez-Gardenes2007,Wu2013,Xu2017}. In this section we will show that the above phenomenon stems from the typical structure that locally compact clusters with sparse inter-connections rather than the hubs.

\begin{figure}
\centering
\scalebox{0.3}{\includegraphics{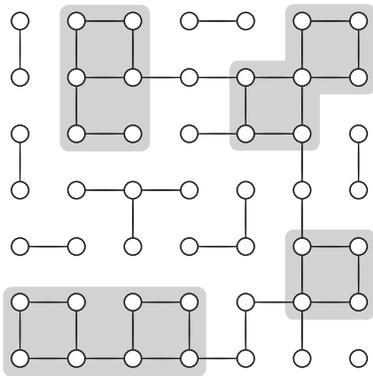}}
\caption{A schematic of the configuration of a square lattice at the percolation threshold. A bond percolation with occupied probability $q$ means that removing each link of the network with probability $1-q$. The percolation threshold is the critical point $q_c$ above which there exists a cluster scaling in the order of system size. For the bond percolation on square lattice shown here, the percolation threshold is $q_c=1/2$. Generally speaking, the configuration around the percolation threshold is composed of some compact clusters (indicated by grey) and some sparse connections between them. At the percolation threshold, it just demonstrates a fractal structure \cite{Stauffer1991}.} \label{fig4}
\end{figure}

It is known that the configuration obtained at the percolation threshold also demonstrates a fractal structure \cite{Stauffer1991}, see Fig.\ref{fig4} as an example. It thus provides a framework to adjust a network between a regular lattice and a fractal. Next, we employ the bond percolation model to further explore the effects of the fractal structure generated by diluting a regular lattice.

Put simply, the percolation configuration can be realized by randomly removing a fraction $1-q$ of links from a network, where $q$ is often called occupied probability. If there exists a giant component of the order of system size, we call the system percolates. The percolation threshold is just the critical point $q_c$, above which the system percolates. For bond percolation on square lattice discussed here, the percolation threshold has the exact solution $q_c=1/2$ \cite{Stauffer1991}.

\begin{figure}
\centering
\scalebox{0.3}{\includegraphics{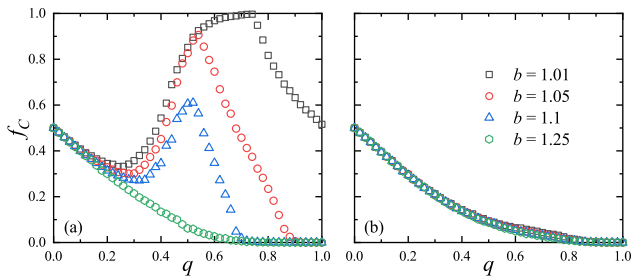}}
\caption{(a) Density of cooperators $f_C$ in diluted square lattices with different occupied probabilities $q$. The network size is $N=10^4$. The cooperation level is optimized at or slightly above the percolation threshold $q_c=1/2$. (b) The case that an individual can only enforce its strategy to the one with a payoff smaller than its own. All the other settings are the same as that used in (a). For this case, the curves for different $b$ almost overlap with each other. All these show that removing some links from the lattice ($q>q_c$) can enhance the cooperation.} \label{fig5}
\end{figure}

As shown in Fig.\ref{fig4}, the configuration around the percolation threshold can also present a typical structure as that shown in Fig.\ref{fig1} (b) but without hubs, i.e., locally compact clusters with some sparse inter-connections. The simulation results shown in Fig.\ref{fig5} indicate that the configuration around the percolation threshold can also enhance cooperation as the fractal structure realized by hierarchical lattices.

Specifically, compared with regular lattices ($q\to1$), the configuration around the percolation threshold can facilitate the emergence of cooperation, and the cooperation level is optimized near the percolation threshold, see Fig.\ref{fig5} (a). If individuals cannot enforce their strategies to the ones with a larger payoff, this optimization disappears, and the cooperator density $f_C$ decreases monotonically with the increasing of $q$, see Fig.\ref{fig5} (b). However, whatever the strategy update rule, Fig.\ref{fig5} suggests that removing some of links from the regular network can enhance the cooperation, i.e., showing a larger cooperator density $f_C$ than that of $q=1$. Note that this makes sense only above or near the percolation threshold. When $q<q_c$, the network falls to pieces \footnote{Strictly speaking, below the percolation threshold, all the clusters are very small and scales in the order of $\log N$. However, for a finite system, a cluster of order $N$ might also exist in the area that not far away from the percolation threshold.}, and the system state is almost entirely determined by the initial distribution of the two strategies.

The main difference between the two fractal structures shown in Figs.\ref{fig1} and \ref{fig4} is that the former has hubs, i.e., the nodes with very large degrees. This suggests that the found manner of the fractal structure does not rely on the existence of hubs. From the fact that the cooperation is significantly enhanced for a wide range of $q$ (optimized near the percolation threshold), we can further know that a standard fractal structure (exactly at the percolation threshold) is not needed to exhibit this phenomenon, as long as the typical structure exists. This can be also seen from the results in the last section, which shows a high cooperator density for a wide range of $p$.

Nevertheless, our finding does not mean that all the fractal structures have such a property. A typical example is the giant component of ER networks at the percolation threshold, i.e., the one with average degree $1$. For this case, the connections of the giant component is so sparse that the locally compact cluster is absent. It is known that the organization and the stability of cooperator clusters are highly dependent on the local connections of individuals \cite{Perc2017}. Therefore, this fractal structure cannot constitute sufficient conditions to sustain cooperation. This is consistent with the results of Ref.\cite{Tang2006}, which suggested that the density of cooperators peaks at a large average degree depending on $b$.

\section{Conclusion}

In this paper we discussed the organization of cooperation in the fractal structure. As we know, in the spatial evolutionary game, cooperators can survive only by forming clusters, while the fractal structure naturally provides many locally compact connections for the forming of cooperator clusters. We found that when $b$ is not too large the sparse inter-connections between these clusters construct many asymmetric barriers for the growth of C and D clusters, for which D strategy is almost impossible to cross, but C strategy has a not too small chance. This suggests that the network with a long average distance could also show a very high cooperation level. In contrast to the fractal structure, there are some works also showed that the asymmetric spreading of strategies can be observed/constructed by introducing different behaviors of individuals \cite{Szolnoki2009,Szolnoki2018,Szolnoki2014}.

We also pointed out that the scale of these compact clusters is much smaller than community. The community is a macrostructure, in which C and D can coexist. Here, the small compact clusters are often fully occupied by either C or D. However, due to sparse inter-connections between clusters/commnuties, both of them lead to a long convergence time.

In summary, we found that the fractal structure does not always play a negative role in the organization of cooperation. Compared with regular networks, the fractal structure can also facilitate the emergence of cooperation, suggesting that both removing and inserting links from/into a regular network can enhance cooperation. This means that a global scale communication channel is not strictly necessary to improve cooperation. By optimizing local connections, the cooperation can also be enhanced. This has important implications for our understanding of the emergence and organization of cooperation in different scales.

\section*{Acknowledgments}
The research of D.P. was supported by the Doctoral Scientific Research Foundation of Anhui University (Grant No. Y040418184).

\bibliography{ref}

\end{document}